\documentclass[11pt]{article}
\usepackage[margin = 1 in]{geometry}
\usepackage{color}
\usepackage{amsmath}
\usepackage{textcomp}
\usepackage{graphicx}
\numberwithin{equation}{section}
\newcommand{\abs}[1]{\left| #1 \right|}

\newcommand{\be}{\begin{equation}}
\newcommand{\bea}{\begin{eqnarray}}
\newcommand{\ee}{\end{equation}}
\newcommand{\eea}{\end{eqnarray}}

\def\A{\mbox{\bf A}}

\def\x{\mbox{{\bf x}}}

\def\X{\mbox{{\bf X}}}

\def\Y{\mbox{$\bf{Y}$}}
\def\y{\mbox{$\bf{y}$}}

\def\p{\mbox{$\bf{p}$}}

\def\f{\mbox{$\bf{f}$}}

\title{{\bf Strategic Monte Carlo Methods\\ for State and Parameter Estimation\\ in High Dimensional Nonlinear Problems}}
\bigskip
\bigskip
\author{Sasha Shirman,\\ Department of Physics, \\ University of California, San Diego\\
and\\
Henry D. I. Abarbanel,\\ 
Department of Physics\\
and\\
(Marine Physical Laboratory) Scripps Institution of Oceanography\\ 
 University of California, San Diego}
\bigskip
\date{\today}

\graphicspath{%
    {converted_graphics/}
    {/}
    {2meas/}
    {4meas/}
    {5meas/}
    {7meas/}
}
\begin{document}
 \bibliographystyle{apalike}
 \raggedbottom
\maketitle
\begin{abstract}
In statistical data assimilation one seeks the largest maximum of the conditional probability distribution $P(\X,\p|\Y)$ of model states, $\X$, and parameters,$\p$, conditioned on observations $\Y$ through minimizing the `action', $A(\X) = -\log P(\X,\p|\Y)$. This determines the dominant contribution to the expected values of functions of $\X$ but does not give information about the structure of $P(\X,\p|\Y)$ away from the maximum. We introduce a Monte Carlo sampling method, called Strategic Monte Carlo (SMC) sampling, for estimating $P(\X, \p|\Y)$ in the neighborhood of its largest maximum to remedy this limitation. SMC begins with a systematic variational annealing (VA) procedure for finding the smallest minimum of $A(\X)$. SMC generates accurate estimates for the mean, standard deviation and other higher moments of $P(\X,\p|\Y)$. Additionally, the random search allows for an understanding of any multimodal structure that may underly the dynamics of the problem. SMC generates a gaussian probability control term based on the paths determined by VA to minimize a cost function $A(\X,\p)$. This probability is sampled during the Monte Carlo search of the cost function to constrain the search to high probability regions of the surface thus substantially reducing the time necessary to sufficiently explore the space.
\end{abstract}

 \newpage

\section{Introduction}

Statistical data assimilation transfers information from observations, performed in a time interval $[t_0,t_1,...,t_F = t_N]; t_n = t_0 + n \Delta t$, to the states and parameters of a model of the processes producing the measurements. The goal is to accurately estimate the unknown parameters and unobserved states. The model has $D$ state variables $x_a(t); \;a=1,2,...,D$ and $N_p$ parameters in the vector $\p$. $L$-dimensional observations $y_l(\tau_k); l=1,2,...,L \le D + N_p$ are made at times $\{\tau_k\}\;;k = 1,2,...,F;\; t_0 \le \tau_1 \le \tau_2 \le ... \le \tau_F \le t_F$ in the measurement window $[t_0,t_N]$. 

The dynamical rule moving the model from time $t_n$ to time $t_{n+1}$ is given by the discrete time map $\x(t_n) \to \x(t_{n+1}) = \f(\x(t_n),\p)$. $L \le D+N_p$, but usually $L \ll D+N_p$ in practice. We denote the collection of model states and parameters as $\{\x(t_0),\x(t_1),...,\x(t_F), \p \} = \{\X,\p\};\;t_0 \le t_n \le t_F, n=0,1,2,...,N$; $\X$ is the {\em path} of the model state variables through the observation window. The collection of observations is designated as $\Y =\{ \y(\tau_1),\y(\tau_2),...,\y(\tau_F)\}$. As the measurements are noisy and the model has errors, we seek the conditional probability distribution $P(\X|\Y)$ over the observation window. For brevity and clarity we often denote the path as just $\X$ supressing the explicit dependence on $\p$ and setting the measurement times, $\tau$, and the model integration times, $t$, to be equal.

Introducing the `action' $A(\X) = -\log [P(\X|\Y)]$~\cite{Action} we may write the conditional expected value of a function $G(\X)$ on the `path' $\X$ as
\be
E[G(\X)|\Y] = \langle G(\X) \rangle = \frac{\int d\X\, G(\X) e^{-A(\footnotesize \X \normalsize)}}{\int d\X\,  e^{-A(\footnotesize \X \normalsize)}}.
\label{expected}
\ee
An interesting function on the path $\X$ is the path itself $G(\X) = \X$.

To perform the integral one may use the method of~\cite{laplaceold,laplacenew} which entails seeking the path $\X^0$ associated with the smallest minimum of the action $A(\X)$. The path $\X^0$ contributes to the expected value integral approximately $\exp[A(\X^1) - A(\X^0)]$ more than the path $\X^1$ associated with the second smallest minimum. One may also use a Monte Carlo method to seek the paths $\X^q;\;q=0,1,2,...$ associated with the minima of the action $A(\X^q)$ and to sample in the regions near the $\X^q$. 

It would seem that expected values may be estimated accurately using only the maxima identified through Laplace's method~\cite{laplaceold}. For probability distributions with steep and narrow peaks this may be the case. A current method developed by~\cite{ye-et-al} called Variational Annealing (VA), deals with the difficulty of maximizing a $P(\X|\Y)$ expressing a complex and nonlinear surface. VA is an optimization method in which the dynamical constraints of the model dynamics are slowly enforced.

Analysis through VA provides no information about the structure of $P(\X|\Y)$ away from its maxima.  In the general case when $P(\X|\Y)$ does not have skinny tails, it is necessary to more fully understand the structure of $P(\X|\Y)$ to accurately estimate the expectation value and higher moments of any function on paths: $\langle G(\X) \rangle$.

To remedy this limitation, this paper melds these two methods through utlizing a variational annealing~\cite{ye2014precision}~\cite{ye2015physrev} approach to finding the $\X^q$ and then strategically sampling near them using a Monte Carlo protocol. This method is very successful in finding the locations in path space of the many local maxima of $P(\X|\Y)$ and additionally locating an apparent global maximum for the probability.

We investigate an extension to VA that allows for informed exploration of the structure of $P(\X|\Y)$. In multimodal systems, we are interested only in the structure of $P(\X|\Y)$ near the largest maxima as those regions contribute most significantly to the expected values of the statistical quantities of interest. 

We develop and describe here a method through which the results of a VA analysis are used to inform the boundaries of a Metropolis-Hastings~\cite{MC} search in path space. Constraining the search to regions of interest as determined by VA allows for reduced computational time to complete a search of sufficient length to estimate the structure of $P(\X|\Y)$. Combining a variational method with a constrained random search provides the benefits of both the faster variational method and the more thorough and informative Monte Carlo search.

A previously developed method, the Maximum Likelihood Ensemble Filter method~\cite{MLEF}, also attempts to provide more accurate information as to the structure and higher moments of the distribution. This method does not address error introduced through inaccuracies in the model. This enforces strong model constraints that may lead to a search surface punctuated by many local minima. The strategic Monte Carlo (SMC) method presented in this paper differs from this in part by allowing for error in the model as addressed in~\cite{Action}~\cite{ye2015physrev}.

We test our SMC
on the Lorenz96~\cite{LChaos2} system with $D = 11$. 
A simulation is used to generate model states and a varying amount of information is presented to the model during the analysis. We search independently in each component of path space leading to a problem with $D(t_F - t_0) + N_p = DN + N_p$ state variables. $N_p$ is the number of parameters in the model. In this paper we project the $D(t_F - t_0) + N_p$- dimensional probability distribution into an $N_p$ dimensional distribution. This is done by integrating out the state variable dependence via $P(\p) = \int d\X P(\X, \p)$.
We expect to see increased sharpness of the local probability maxima in path space and a narrowing of $P(\X|\Y)$ as the amount of information in the data is increased. As the amount of observed data becomes quite large, we expect $P(\p)$ to approach a delta function. Our results indicate this to be the case.

\section{Background}


We have written the conditional probability density function $P(\X,\p|\Y)$ as:
\small
\bea
 &&P(\X, \p | \Y)  \propto e^{- A(\footnotesize \X \normalsize, \p)}  \nonumber \\
 &&A(\X, \p) = \sum_{t=0}^{t_F}  \biggl[ \sum_{a = 1}^{L}  \frac{R_m(a,t)}{2} \left(x_a(t) - y_a(t)\right)^2 +\sum_{a = 1}^{D} \frac{R_f(a)}{2}(x_a(t+1) - f_a(x(t),\p))^2 \biggr ],
\label{standard}
\eea
\normalsize
in which $R_m(a,t)$ is the (diagonal) precision matrix for the Gaussian noise in the measurements, and $R_f(a)$ is the (diagonal) precision matrix for the Gaussian error in our model $\x(t+1) = \f(\x(t),\p)$. 

The first term in $A(\X, \p)$ in Eq. (\ref{standard}) reflects a Gaussian error in the measurements, while the second term describes the model error and defining the correlations between variables, also taken to be Gaussian. $P(\X, \p | \Y)$ is not Gaussian as $\f(\x,\p)$ is nonlinear. While other choices of error distribution are certainly possible, this expression of measurement error and model error is analyzed so frequently we call this the `standard model' for statistical data assimilation. The expected value integrals of interest, Eq. (\ref{expected}), are not Gaussian as the function $\f(\x,\p)$ is nonlinear in the state variables of the model.

Other methods of DA, such as various modifications of Kalman filtering, typically use {\em ad hoc} definitions of variable correlations in the formulation of the cost function. Though informed by the system and dynamics, these correlation matrices are forced to certain values~\cite{EKF}. In contrast, the methods in this paper define the variable correlations in the cost function through the model dynamics which specify both spatial and temporal coupling of states.

When $R_f \to \infty$ the model is satisfied exactly (becomes deterministic). Alternatively, when $R_f \to 0$, the standard deviation of model error is infinite and nothing can be learned about the time evolution of the system. At $R_f = 0$ the probability distribution is a Gaussian with width given by the measurement noise in the measured state variable directions and flat in the direction of unmeasured variables. When there is zero precision in the model, information is not transfered from measurements to umeasured model states. 

If the discrete time model dynamics defined by the function $f(\x,\p)$ perfectly describes the system, then the probability will approach an approximation of a delta function with width given by the measurement accuracy.
As the number of measurements decreases, the probability will broaden in the unmeasured directions as well as possibly become more complex and multimodal depending on the structure of the equations of motion and the form of the coupling between measured and unmeasured degrees of freedom in $\f(\x,\p)$.


We now turn to a method for using the hyper-parameter $R_f$ as a tool for identifying the smallest minimum of the action. When the smallest of these minima is much smaller than the magnitude of the next minimum, it dominates the expected value integral Eq. (\ref{expected}). Our focus will then be on methods to sample $P(\X,\p|\Y)$ in the region of path space near the path giving the smallest minimum of the action.


\subsection{Variational Annealing (VA)}
In VA, as in other optimization methods, the maximum of the conditional probability $P(\X, \p|\Y)$ is sought. 
It is assumed that the expectation value of a function on the path $\X$ can be approximated using this maximum.
This relies on a global maximum with skinny tails so it is significantly larger in magnitude than other local maxima \cite{laplaceold}. 
Estimating the global maximum of $P(\X, \p|\Y)$ or equivalently the global minimum of the action $A(\X, \p) = - \log[P(\X, \p|\Y)]$ for actions nonlinear in $\X$  is an NP-complete problem~\cite{murty87}, and, for practical purposes is numerically intractable, unless there is a special circumstance. In our action, Eq. (\ref{standard}), we have a special circumstance.

VA is an iterative method through which model constraints on a variational principle, such as those imposed by the equations of motion, are gradually enforced through increases in $R_f$ from $R_f \approx 0$. At the first iteration of the VA procedure, the model is unenforced by setting $R_f = 0$. This gives a poor approximation for the probability density which is Gaussian in the measured directions and uniform in the unmeasured ones. We slowly increase the value of $R_f$ in such a way that the structure in path space, $\X$ of the probability density $P(\X|\Y)$, changes adiabatically between iterations. We use the location of the maxima from one iteration of the process as the initialization of the optimization routine at the next iteration of the procedure. 

$R_f$ is changed through the rule
\begin{equation}
 R_f = R_{f0} \alpha ^\beta,
\end{equation}
where $\alpha > 1$ and $\beta = 0, 1, 2, ...$.

This process assumes that, with small enough $\alpha$, maxima from one iteration will move somewhere close to the peak of the maxima in the next iteration. This allows VA methods to `track' the change in probability maxima, action minima, as $R_f$ slowly increases.

When $R_f$ is sufficiently large, the annealing procedure may be terminated and the location in path space of several local maxima may be estimated. 
The $R_f$ value with the best prediction for $t > t_F$, as compared to a continuation of the measured data, is chosen as the `correct' $R_f$ value. This value of $R_f$ is used to calculate the desired probability in Eq. (\ref{standard}) that used in the Monte Carlo search described in the next section.


\subsection{Importance Sampling Monte Carlo}

A Monte Carlo procedure is a type of random walk search. In these searches, a `walker' moves through path space with the probability landscape guiding the direction it steps. Over a long enough walk, the amount of time spent in a region of state space becomes proportional to the probability of those states. When sampling from a distribution using a Metropolis-Hastings algorithm, first developed by~\cite{MC}. The search procedure consists of two steps:

\begin{enumerate}
  \item start at some point in path space $\{\X,\p\}$; choose, or sample, a new point $\{\X',\p'\}$ in path space selecting from $P_{sample}(\X', \p')$
 \item decide whether or not to {\bf accept} this new point and to `step' to it with $P_{accept}(\X', \p')$, or
{\bf choose to reject} this new point, and repeat the whole procedure starting with the same initial point in $\left\{X, \p \right\}$.
\end{enumerate}

It is important that the distribution from which the `new' point is sampled overlaps strongly with the search distribution. Intuitively this is because it is impossible to search a region that is unsampled. Mathematically, this is because the distribution that is searched is equal to the product of the sampling distribution and the acceptance/rejection distribution. 

If the sampling distribution is zero where sampling occurs, the search will result in a zero probability regardless of the acceptance/rejection probability since
\be
P_{output} (\X, \p|\Y) = P_{sample}(\X, \p|\Y) \cdot  P_{accept}(\X, \p|\Y).
\ee
For notational simplicity we drop the dependence on the data, $\Y$ in the following discussion.

We initialize our walker at location, $(\X, \p)$, and sample a location, $(\X', \p')$. The probabilities of sampling the next location and of accepting it are conditioned on the previous walker location. We use the property of detailed balance which states that
\begin{equation}
 \frac{P(\X', \p')}{P(\X, \p)} = \frac{P(\X', \p'|\X, \p)}{P(\X, \p|\X', \p')}
\end{equation}
to connect the absolute probability of a state with a conditional probability of a state of the walker given its previous state.

Then we have
\begin{equation} \label{probs}
 \begin{gathered}
\frac{P(\X', \p')}{P(\X, \p)} = \frac{P_{sample}(\X', \p'|\X, \p) P_{accept}(\X', \p'|\X, \p)}{P_{sample}(\X, \p|\X', \p') P_{accept}(\X, \p|\X', \p')}
 \end{gathered}
\end{equation}
A common, and easy to sample, choice for these probabilities is $P_{sample}(\X', \p'|\X, \p)$ selected as a Gaussian centered at $\X,\p$ with some standard deviation given by the desired `step size', where $P_{sample}$ is a function of $|\X-\X'|$.

A sampling probability that is symmetric transforms Eq. (\ref{probs}) to the simplified expression:
\begin{equation}
 \frac{P(\X', \p')}{P(\X, \p)} = \frac{P_{accept}(\X', \p'|\X,\p)}{P_{accept}(\X, \p|\X', \p')}
\end{equation}

We then can set 
\begin{equation} \label{accept} P_{accept}(\X', \p'|\X, \p) = \min\left[1,\frac{P(\X', \p')}{P(\X, \p)}\right]
\end{equation} 
as a possible acceptance probability. These choices uniformly sample the state space of the problem. The search is allowed to run for a sufficiently long time that the location of the walker becomes decorrelated from the initial location. The probability of a location is subsequently determined by the fraction of time the walker spent at this location.

If some structure of the desired distribution is known, it is possible to explore only the higher probability regions of state space using the methods of Importance Sampling as described in \cite{ImportanceSamplingChapter}. A sampling probability $P_{sample}(\X', \p'|\X, \p)$ is chosen such that it depends explicitly on $\X$ as well as the distance between $\X$ and $\X'$. This sampling probability is often smoother and easier to sample than the desired distribution, but it gives a rough approximation of the target distribution. For example, 
\begin{equation}
 P_{sample}(\X'|\X) = P_{bias}(\X', \p) P_{step}(\X', \p'|\X, \p)
\end{equation}

Where $P_{step}$ is the previous $P_{sample}$, a Gaussian with standard deviation given by desired step size and symmetric in $\X$ and $\X'$. The acceptance probability is left unchanged. Consequently, the output distribution is given by 
\begin{equation}
 P_{output}(\X', \p') = P(\X'|\p') P_{bias}(\X', \p')
\end{equation}

To estimate the desired distribution of $P(\X', \p')$, the output of the search algorithm is multiplied by a weight given by 
\begin{equation}
W(\X', \p') = \frac{1}{P_{bias}(\X', \p'|\X, \p)}.
\end{equation} 
This weight removes any bias introduced into the outcome by sampling certain regions more often than others. To thoroughly explore all regions of interest or high probability, it is necessary that $P_{bias}(\X', \p')$ overlap strongly with the high likelihood regions of $P(\X', \p')$.

\section{Methods}
As stated previously, maximization of the probability distribution through the optimization of the action provides a good estimate for the location of many of the maxima of the conditional proability density function. Additionally, through the use of VA we produce a reasonable estimate of the global maximum of this density. In many cases, such as those in which the probability is asymmetric or has fat tails, the location in state space of the maxima is insufficient to estimate expectation values and higher moments for a distribution. An approximation of the structure of the probability is needed. 

We implement an informed random walk that is constrained to regions of interest in state space by the results of the previous VA analysis. In other words, we execute an importance sampling Monte Carlo in which the sampling distribution is informed by the location of the local maxima in the probability as determined by the variational annealing method.

Prior knowledge of the desired distribution acquired through optimization can be used to determine an appropriate bias probability, $P_{bias}$, such that the bias probability is approximately centered at the maximum probability region of the desired probability density and wide enough to encompass all regions of interest.

As the search space, and thus the resulting probability density function, is extremely high dimensional, it is useful for visualization of the results to project the estimated distribution into a lower dimensional space. Because parameters are constants in motion and because they complete describe the model dynamical system, when the functional form of $\x(n+1) = \f(\x(n),\p)$ is given, we project the probability density function to the smaller space of the parameter vector.

\begin{equation}
\begin{gathered} \label{normalization}
 P(\p') = \int {P_{output}(\X', \p')} W(\X', \p') d\X'
 \end{gathered}
\end{equation}

The integral in \ref{normalization} is estimated as a sum over all accepted steps in the walker's path. In the case of the Lorenz96 system with only one parameter, this sum produces a probability distribution in one dimension which can be easily visualized.

\subsection{Informing the Random Walk through VA} \label{choosing}
To best utilize the information aquired through VA, we use the location of the probability maxima determined to define the $P_{bias}$ used to constrain the random walk. We set the center of $P_{bias}(\X', \p')$ at the location of the highest probability path found by the VA method. As a first order approximation, we set the bias probability to be an uncorrelated multivariate Gaussian in all dimensions of $(\X', \p')$. In each dimension the standard deviation of the Gaussian is set to be four times the distance between the location in state space of the largest and smallest local maxima of the probability found by the VA method.

In other words, define a path $(\X_0, \p_0)$ which sets the location of the largest local maximum and $(\X_1, \p_1)$ that sets the smallest local maximum found by VA. Recall that $\X \equiv \left\{ \x(t_0), \x(t_1), \dots, \x(t_F) \right\}$. If only one local maximum is found at the chosen $R_f$, or model precision, value we set $(\X_1, \p_1)$ to be equal to a smaller local probability maximum found at a previous set in the annealing procedure. We define the distance, $D_s$, between these paths as

\begin{equation}
 \begin{aligned}
D_s =&   (\abs{\X_0-\X_1}, \abs{\p_0 - \p_1}) \\
D_s = &\Big(\left\{ \abs{\x_0(t_0) - \x_1(t_0)}, \abs{\x_0 (t_1) - \x_1(t_1)}, \right. \nonumber\\& \left.\left. \dots, \abs{\x_0(t_F) - \x_1 (t_F)} \right\}, \abs{\p_0 - \p_1 } \right) 
 \end{aligned}
\end{equation} 

Where $\abs{\A} \equiv \abs{A_1} \cdot \hat e_1 + \abs{A_2} \cdot \hat e_2 + \dots$.

The dimension of each $\x(t)$ is $D$, the number of dynamical variables. To allow for the same range of search at every time point for a given dynamical variable, we set the standard deviation for the bias probability of the $i$'th element of $\x$ at any time, $t_j$, to be equal to four times the largest distance in that component of $\x$. The factor of four is chosen to maintain weak bias in the sampling while allowing all local probability maxima to fall within the well sampled region of $\{ \X, \p \}$. Or in other words

\begin{equation}
 \sigma_{search}^i (t_j) =  \sigma_{search}^i= 4 \max_{t}{\abs{x_0^i(t) - x_1^i(t)}}
\end{equation}

for the search in state space, and

\begin{equation}
 \boldsymbol{\sigma}_{search}^{p} = 4 \abs{\p_0 - \p_1}
\end{equation}

for the search in parameter space.

The acceptance probability is defined by using Eq. (\ref{standard}) at the current and proposed locations in state space, $(\X, \p)$ and $(\X', \p')$ , in \ref{accept}.

This bias probability which contains all the local maxima found by VA within one fourth of a single standard deviation gently constrains the walkers to stay within regions of high probability as found by VA while still given freedom to explore the structure of the tails of the probability distribution.

 
 \section{Results}
 
 We demonstrate the Monte Carlo Extension to VA in the following section. Testing of this method is done on the Lorenz96 system. The system itself can be characterized by a set of $D$ coupled differential equations with $D \geq 3$. In addition to the dynamical variables, $x_i(t)$ the system has a constant forcing term $G$.
 
 \begin{equation} \label{lorenz}
  \frac{d x_i(t)}{d t} = (x_{i+1}(t) - x_{i-2}(t))x_{i-1}(t) - x_i(t) +G
 \end{equation}
 
The subscripts in $x_i(t)$ are calculated $\mod(D)$. As the magnitude of $G$ grows it moves the behavior of the system from the non-chaotic to chaotic regimes. The transition into chaotic regime occurs at approximately $G=8$ \cite{LChaos1, LChaos2}.

  \subsection{Twin Experiments}
  
 Twin experiments are experiments run on simulated data. These can be used to test methods, design experiments, or determine various requirements such as noise level or measurement frequency when recording data in a real experiment. Twin experiments are useful because having generated the data we know all of the `unmeasured' quantities in the system. This allows us to test whether any estimates of unmeasured states or parameters match those not presented to the analysis and thus whether the data presented provides sufficient information regarding the dynamics to fully describe the system.
  
  A mathematical model, identical to the one that best describes the system of study, is used to generate simulated data. `Noise' is added to the system output to mimic measurement error, with the level of noise matching the known accuracy of the equipment used in a real experiment. A noisy time series of $L\leq D$ variables is presented to the estimation procedure as the only available `data'. Analysis is then done on the simulated data and results are compared with the known underlying dynamics of the system.
  
 For the work discussed in this paper, we dealt with a set of twin experiment examples of the Lorenz96 system. The data for these experiments were collected from a $D=11$ dimensional example with an estimation window containing 165 time points separated by $\Delta t = 0.025 \mbox{ units}$. We ran parameter and state estimations using a varying number $L$ of measured states. And in all cases we added approximately 3.4\% uniformly distributed measurement noise to the `measured' variables. Our experiments varied the number of measured states to demonstrate the dependence of results on $L$.

 \subsection{Estimation on an Eleven Dimensional System}
\begin{figure}[h!] 
  \centering
  \includegraphics[width=\textwidth]{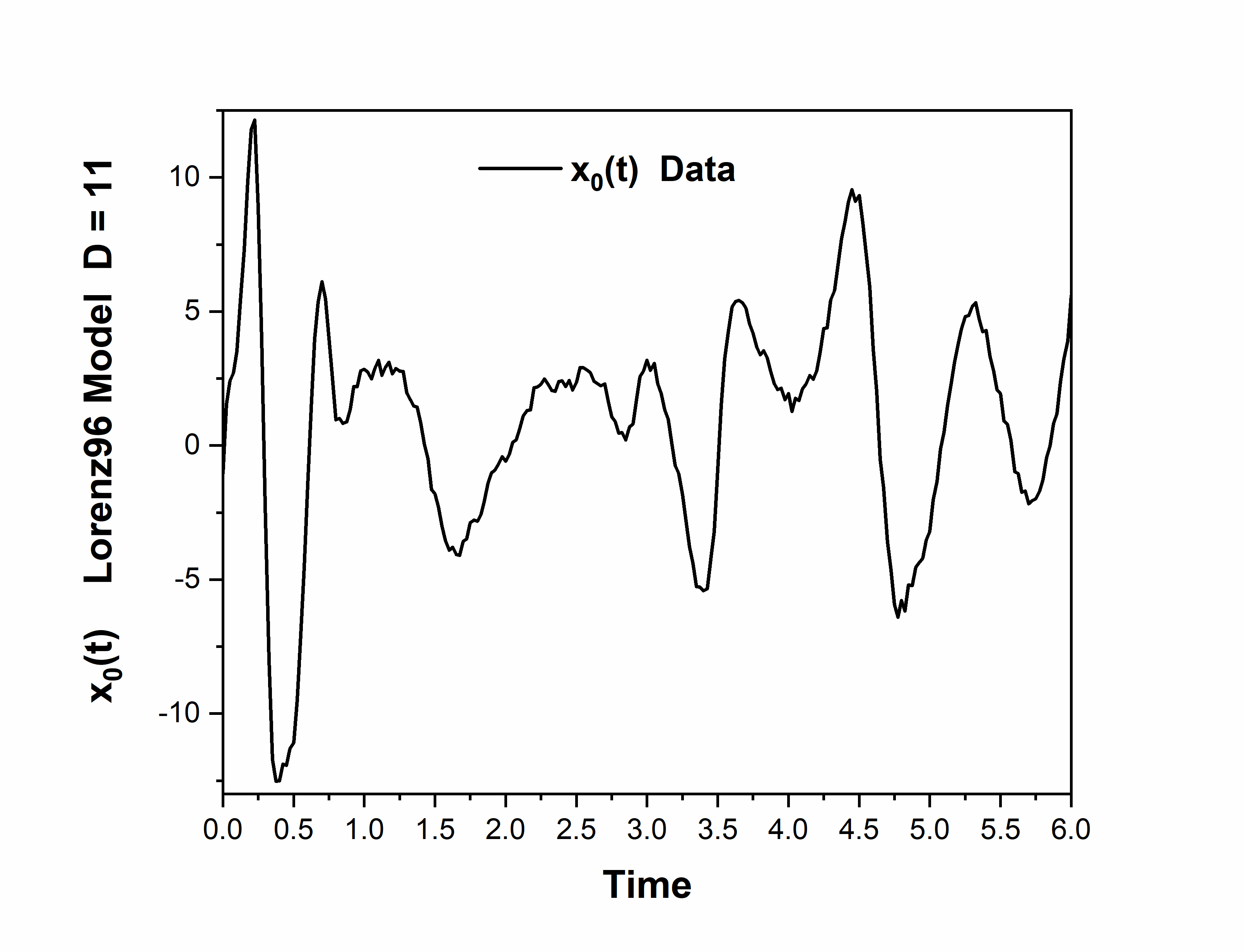}
  \caption{ Time series data for $x_0(t)$ from the Lorenz96 equations $D = 11$. The forcing parameter is $G = 10.0.$}
  \label{timeseries}
\end{figure}

 We set the magnitude of the forcing term in this twin experiment to be $G=10$. Initial states used to generate the data are randomly chosen within the dynamical range of the variables $\approx \left[-10,10\right]$. The equations of motion Eq.(\ref{lorenz}) were integrated forward using the odeInt package in Python. Uniformly distributed noise with magnitude 3.4\% of the dynamical range of the variables of motion was added following the integration. Below is a time series demonstrating the behavior of the first state, $x_0$. This time series shown in Fig. (\ref{timeseries}) is similar to the time series of all other states.
 
 Analysis was done using 2, 4, 5, and 7 measured states. In all cases the measured states were distributed over the system as symmetrically as possible. For example, in the case of two measured states, we measured $x_0(t)$ and $x_5(t)$. This was done because the coupling in Lorenz96 is a type of nearest neighbor coupling and clustering the measured states together does not provide sufficient information to make accurate estimates of unmeasured states~\cite{kostuk}.

 \subsubsection{Variational Annealing}
 
\begin{figure}[h!]
 \centering
  \includegraphics[width=0.45\textwidth]{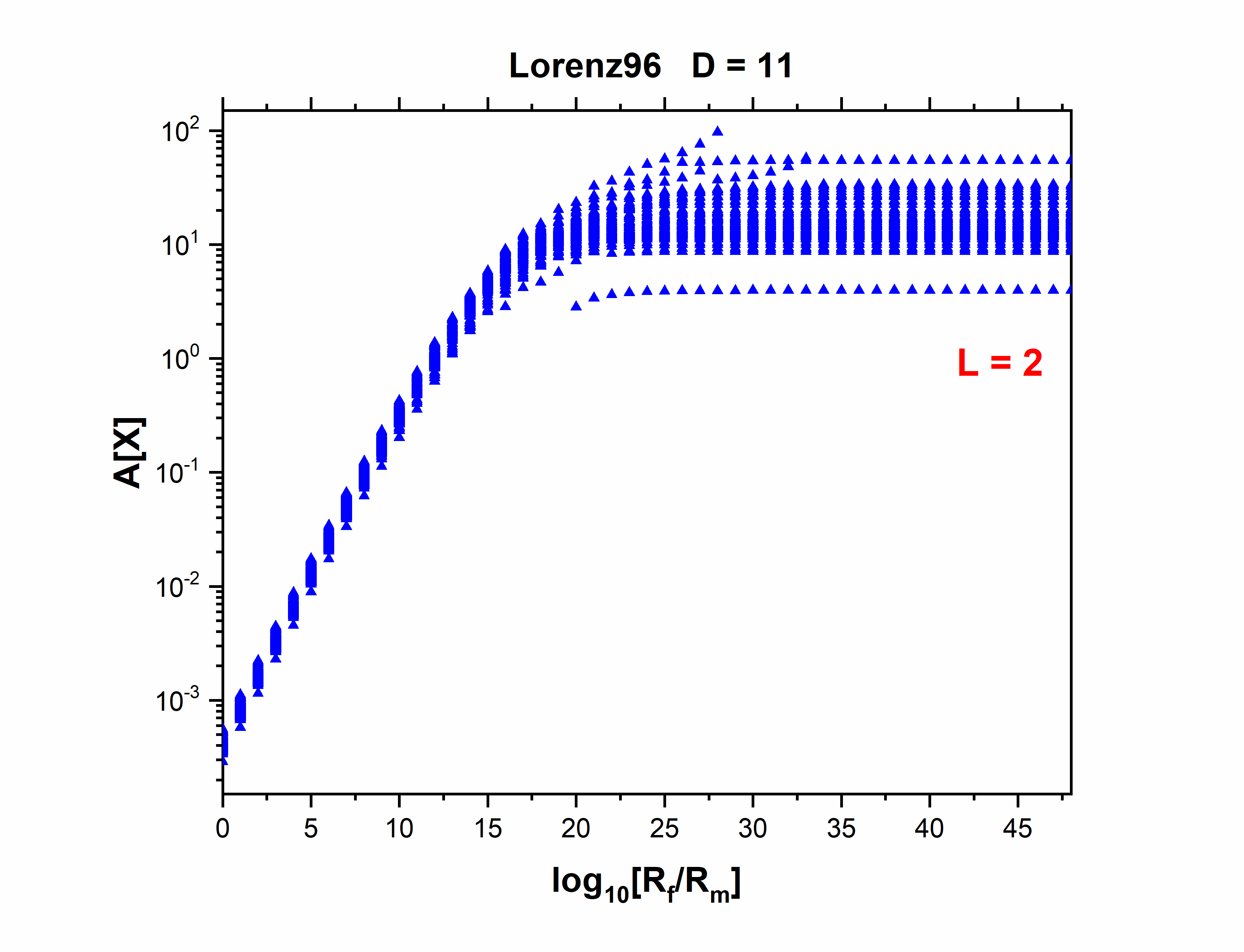}
  \includegraphics[width=0.45\textwidth]{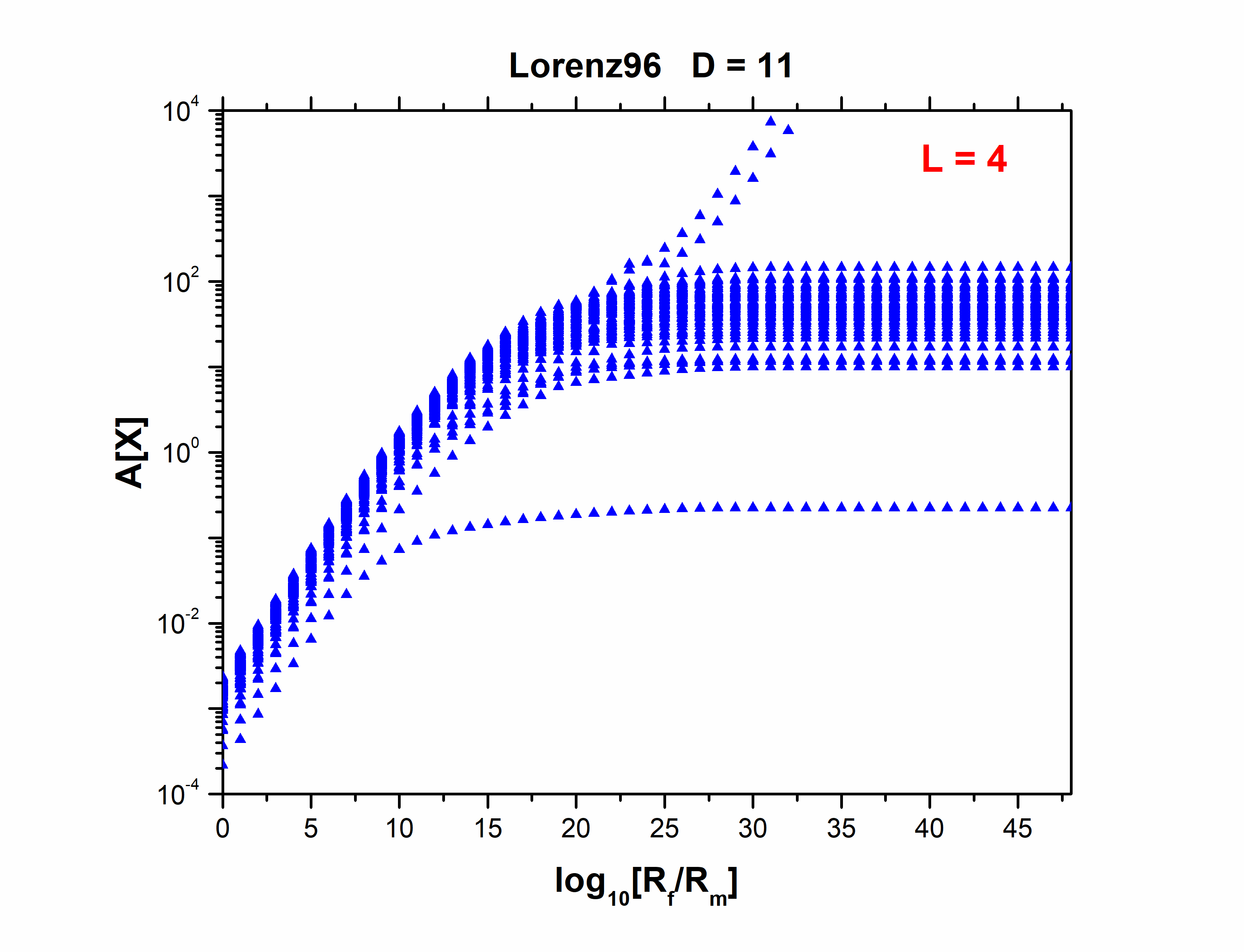}
  \includegraphics[width=0.45\textwidth]{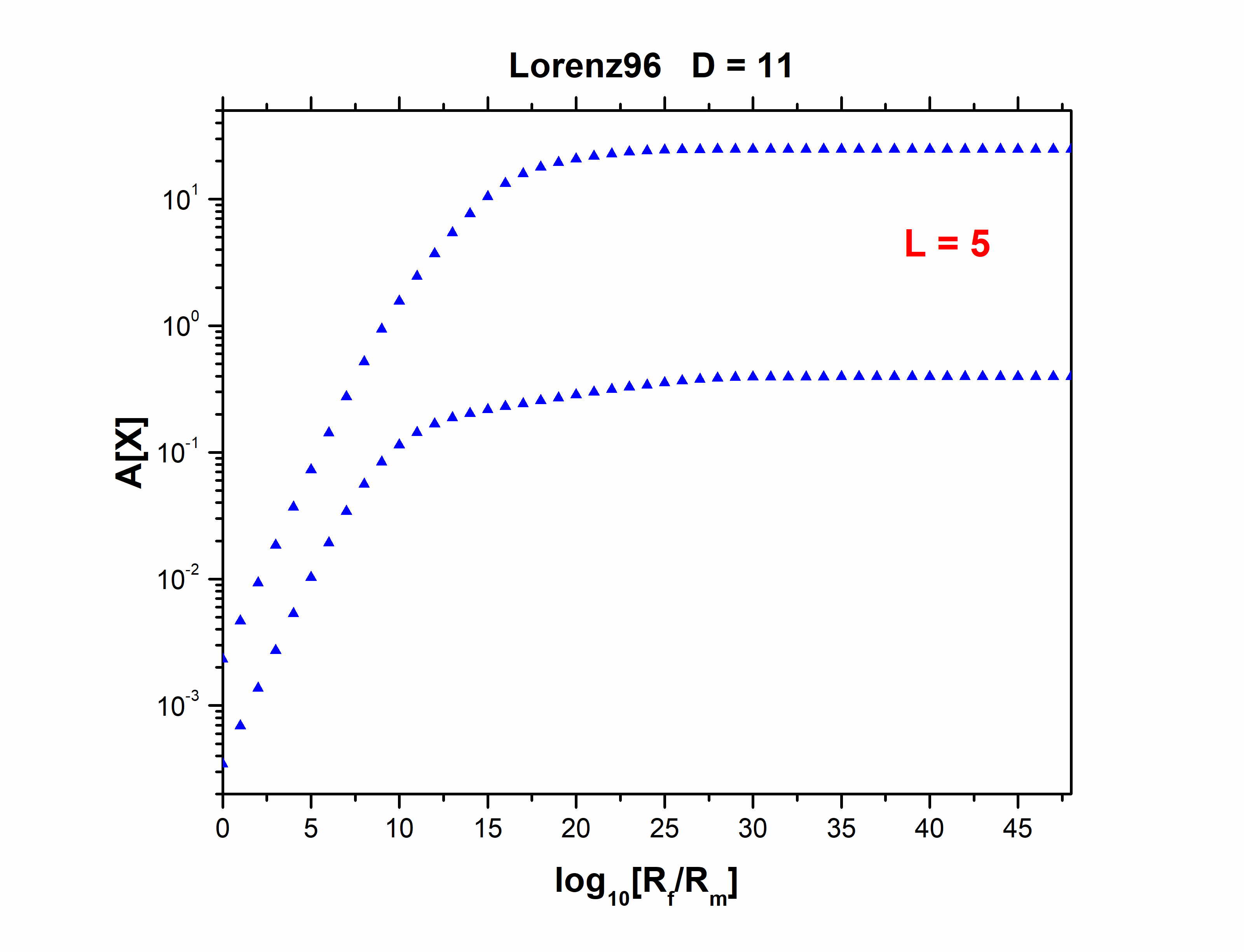}
  \includegraphics[width=0.45\textwidth]{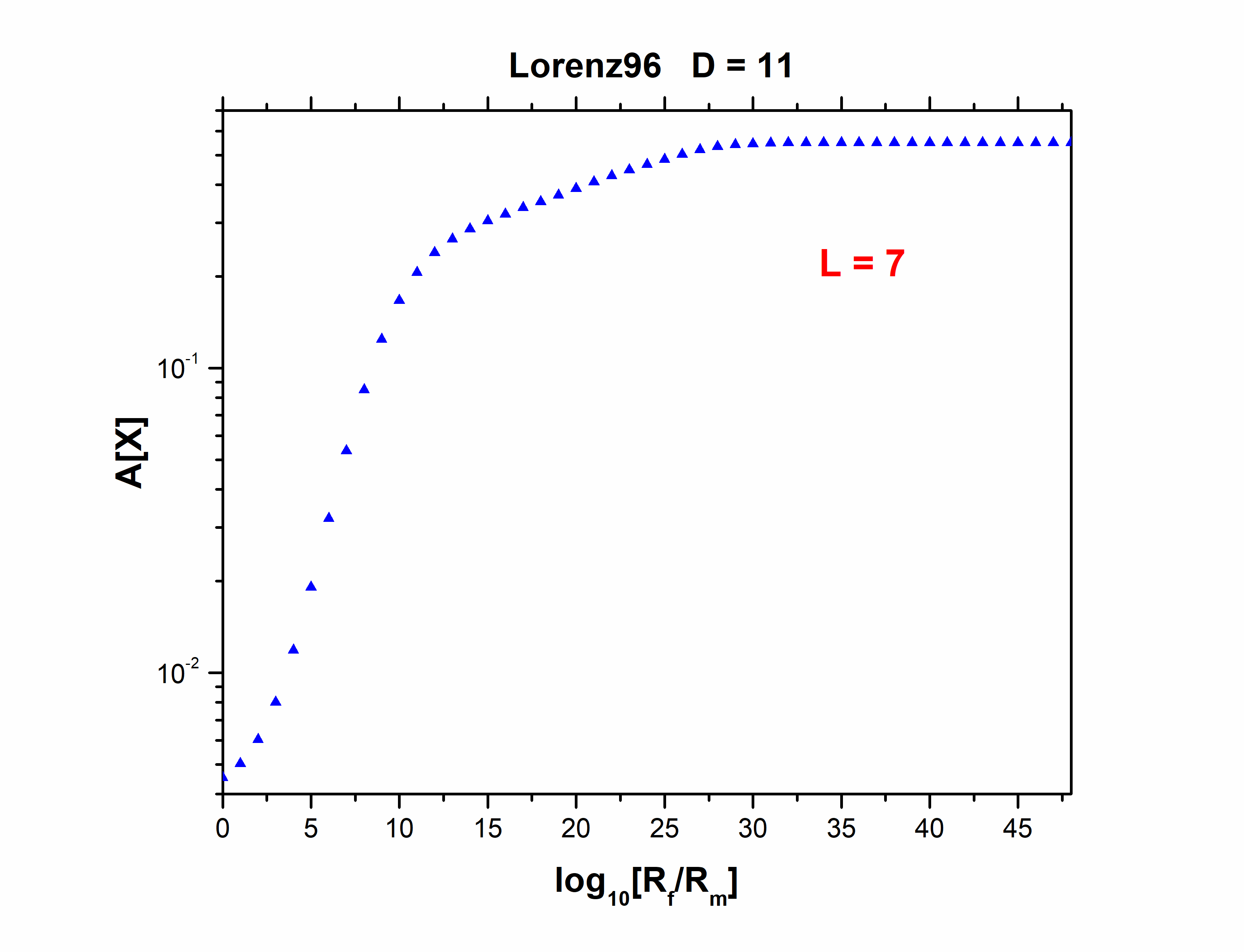}
\caption{Action level plots for the Lorenz 96 model, Eq. (\ref{lorenz}) D = 11, and with L = 2,4,5,7 observed state variables. The number of minima in the standard action Eq. (\ref{standard}) decreases as the number of measurements increase. This implies a smoothing of the action $A(\X)$ and of the probability density function ($\propto \exp[-A(X)]$ with increased information}
\label{actionlor96d11}
 \end{figure}

 Fig. (\ref{actionlor96d11}) demonstrates the results achieved by applying VA to the Lorenz `96 system with L = 2, 4, 5, and 7 measurements respectively. The action is given by the negative log of the probability, $A(\X, \p) = - \log \left[ P(\X, \p)\right]$.  As $\log \left[ \frac{R_f}{R_m}\right]$ increases, the action is increasingly dominated by the model constraint due to the shape of the action given by Eq. (\ref{standard}). Fig. (\ref{actionlor96d11}) shows that as $L$ increases, we see the lowest action/highest probability path increasingly separate from the other local minima. This implies that the probability maximum is exponentially higher on this high likelihood path and that the estimate for the expectation value of the states and parameter will become increasingly dominated by this path. This increasing separation implies that more measured data allows a more dominant maximum likelihood path than when there are fewer measured states. Intuitively this should be clear.
 
 The minimization that produces the plots in Fig. (\ref{actionlor96d11}) was performed using the open source software, IPOPT \cite{ipopt}. The optimization was carried out using 100 initial conditions with 165 time steps of data for each of the plots shown.

 The prediction error was then calculated for every inital condition at every value of $R_f$. Prediction was made by using Python's odeint and integrating forward $T_{pred} = 100$ time steps from the value of the state $\{\x(t_F),\p\}$. The predicted path was compared to the `data' using a least squares algorithm:
 
 \begin{equation}
  MSE = \frac{1}{T_{pred} L} \sum_{t = 0}^{T_{pred}} \left|\y(t_F+t) - \x(t_F +t)\right|^2 \label{error}
 \end{equation}
 
 The squared term in Eq. (\ref{error}) is the vector inner product. Prediction error, $MSE$, was averaged over the number of measurements made so a comparison can be made between the error at each $L$ value. The initialization and $\frac{R_f}{R_m}$ value that generated the lowest prediction error was used to set the location of the center of the Monte Carlo search in path space and the estimate for the value of $R_f$ (the model precision).The initialization and $\frac{R_f}{R_m}$ value that generated the highest prediction error was used to set the width of the sampling probability in path space as discussed in Section \ref{choosing}.

\begin{figure}[htbp] 
  \centering
  \includegraphics[width=0.45\textwidth]{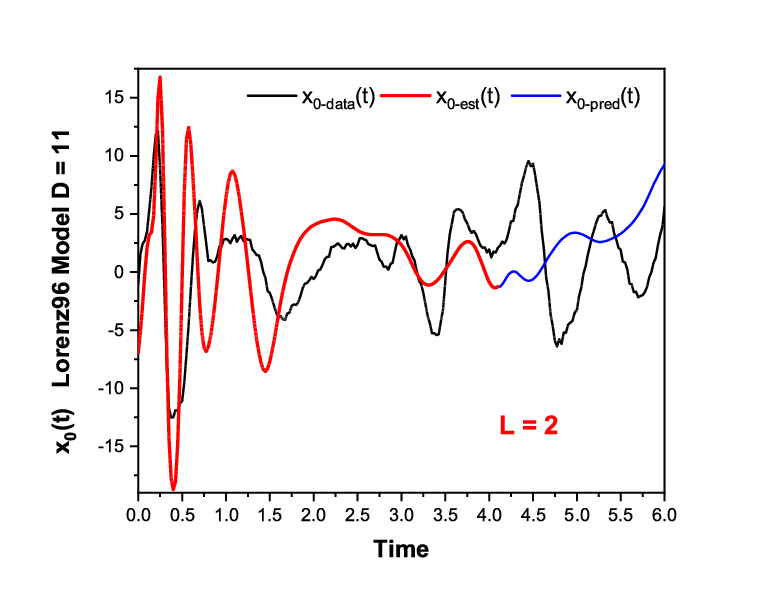}
  \includegraphics[width=0.45\textwidth]{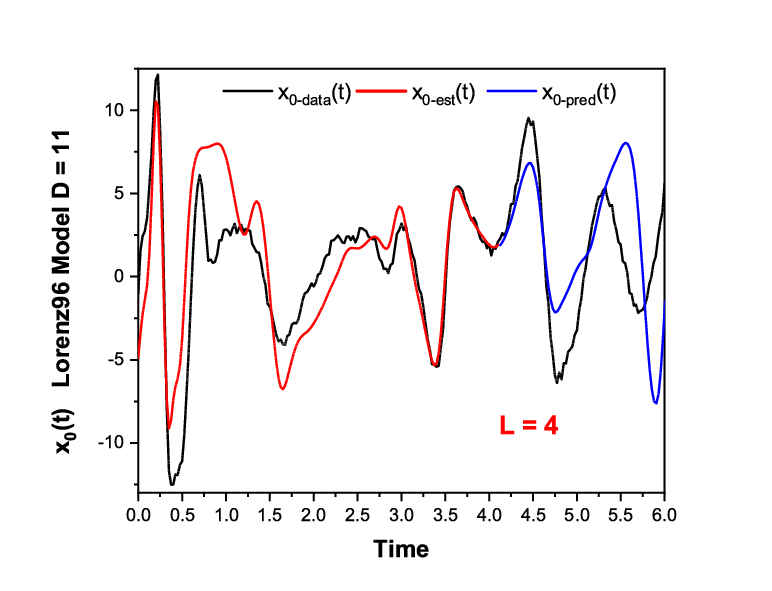}
  \includegraphics[width=0.45\textwidth]{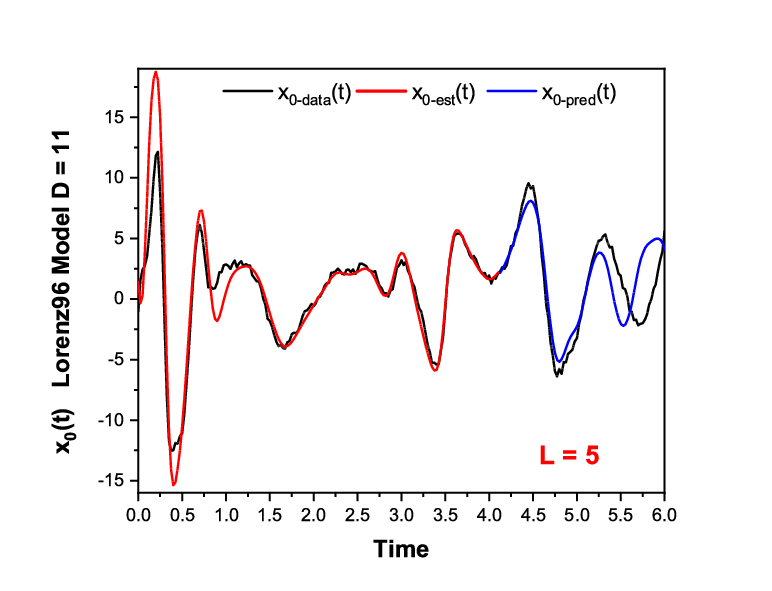}
  \includegraphics[width=0.45\textwidth]{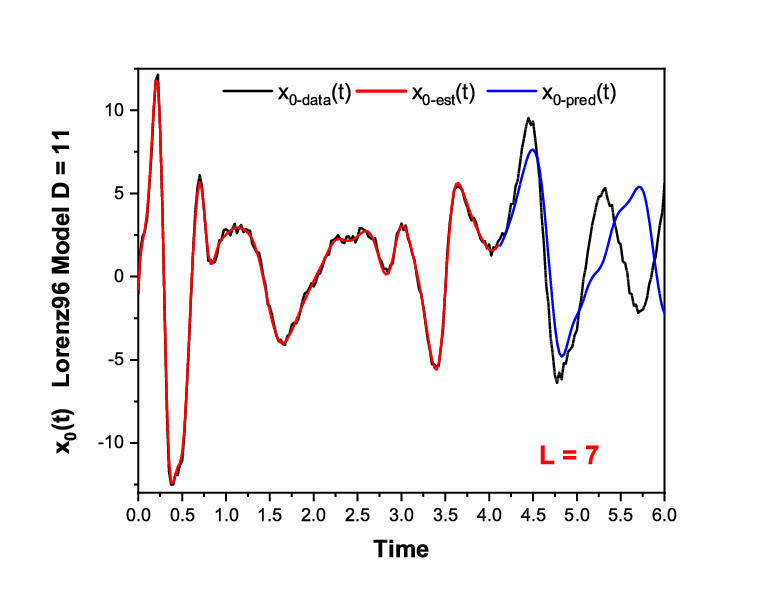}
  \caption{$x_0(t)$ data, estimated, and predicted D = 11, L = 2,4,5,7 measured variables as indicated in red in each figure. 100 prediction time points past the estimation window. SMC is used to estimate 
$x_0(t)$ in the observation window and predict beyond that.}
  \label{predlor96d11smc}
\end{figure}

  \subsubsection{Random Walk}
   
   To estimate the distribution, $P(\p)$, a random walk was initialized with 100 walkers starting at points uniformly distributed over an area of path and parameter space with side lengths equal to $1/2$ the bias standard deviation as defined in Section \ref{choosing}. The random walk was run in Python with a burn in period of length 2,000 steps followed by a 1,000,000 step walk. The location of the walkers in path and parameter space was recorded every 80 steps.


\begin{figure}[htbp] 
  \centering
  \includegraphics[width=\textwidth]{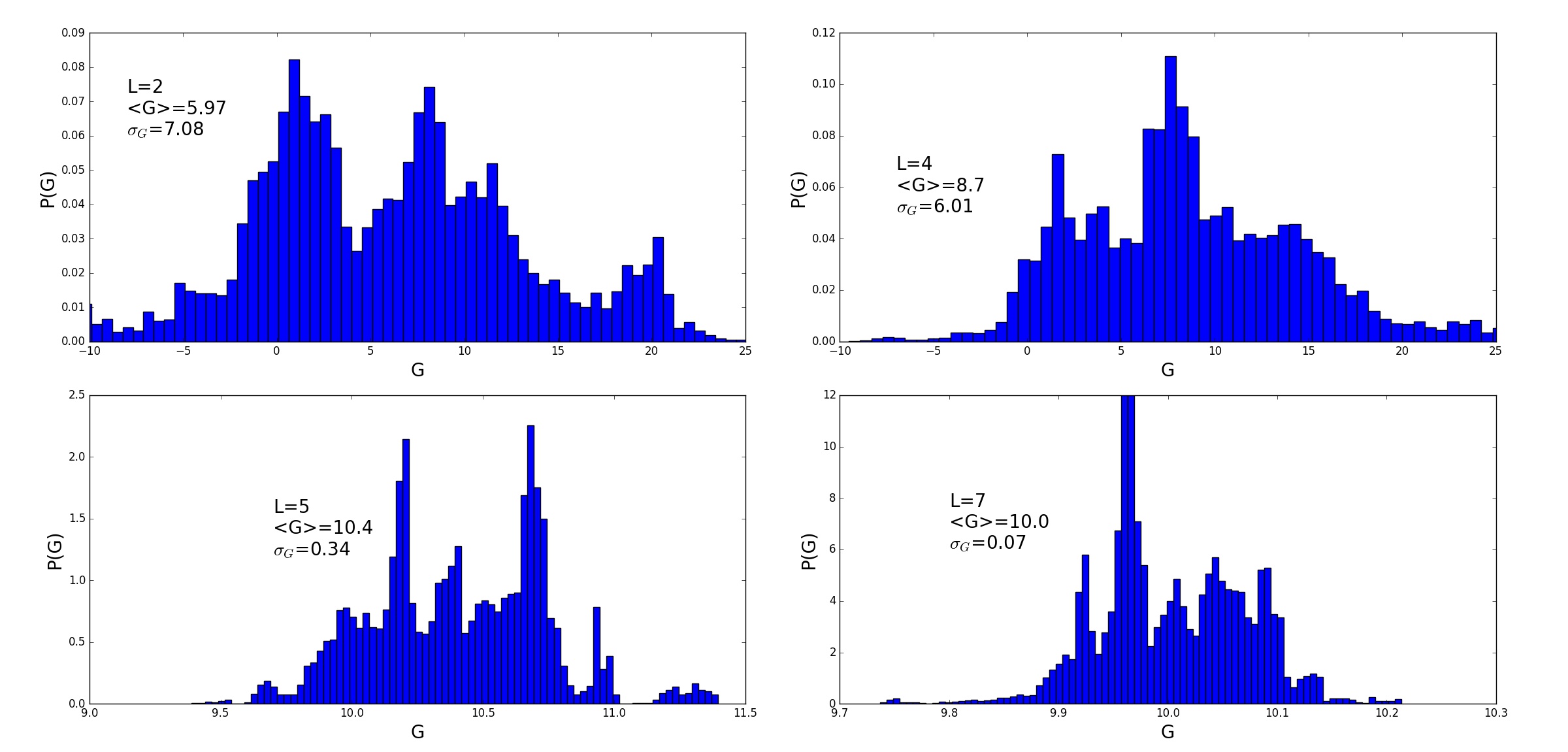}
\caption{Distribution of accepted walker locations as approximation of the underlying probability density function $P(G)$ of the forcing parameter $G$ in the Lorenz96, D = 11 equations. We show $P(G)$ for L = 2,4,5,7. $L$ is shown in each panel in red. The mean value of the distribution $P(G)$ of the forcing parameter $G$ is shown along with the RMS variation around this mean. As the number of measurements $L$ increases, the distributions has a more and more accurate mean for $G$ and becomes narrower and narrower.}
  \label{histlor96d11}
\end{figure}

   Fig. (\ref{histlor96d11}) displays the output of the SMC protocol projected into a histogram $P(G)$ in one dimension for the forcing parameter $G$ as described in \ref{normalization}.
 The histograms in Fig. (\ref{histlor96d11}) are qualitatively what is expected. The bottom right plot in Fig. (\ref{histlor96d11}) demonstrates multiple peaks in probability for 7 measured states with the largest one being much larger than the others. Additionally the width of the distribution tightens as the number of measured states increases. The estimate of the expectation value of the forcing parameter, found in the upper left corner of the histograms, also improves with increased $L$. This is as it should be as an increase in the number of measured states $L$ implies an increase in the amount of information presented to the calculation. This, in turn, allows for better estimates of unmeasured quantities.

We have evaluated these histograms $P(G)$ for various $L$ with a selection of intial locations in path space for the walkers. While the details of the distribution $P(G)$ are different in each case, the mean value is 10.0 within 1\% and the RMS variation about the mean falls in the range $[0.059,0.083]$ for our few examples at $L = 7$.

  \section{Discussion}
We have shown that using random walk methods  to sample conditional probability distributions in the neighborhood of their maximum, informed by our variational annealing numerical optimization to locate that maximum~\cite{ye2014precision,ye2015physrev}, provides useful information as to the structure of the probability distribution. Understanding this structure gives us information as to accuracy of parameter and unmeasured state estimates, which can be useful in applications where a confidence interval is desired. Additionally, the exploration of the structure of conditional probability densities confirms previous assumptions of the dependence of the probability distribution on the number of measured variables.
  
  
The current method demonstrates the expected smoothing of the probability density function as the amount of information is increased and demonstrates promise in approximating higher order estimates of moments of the distribution given the multimodal structure of $P(\X,\p)$. 
The numerical output in the form of estimated mean and standard deviation of the $P(G)$ distribution appears to be relatively stable. Further analysis must be done to calculate the variation in the estimated mean and standard deviation for larger amounts of recorded data, but preliminary results appear to be promising. Regardless of the exact search method used, SMC demonstrates a informed method through which the search space of a high dimensional and multimodal probability may be constrained to only regions of high likelihood which dominate the integral used to calculated expected values.
  \clearpage

\bibliography{mybib}

\begin{thebibliography}{}

\bibitem[Abarbanel, 2013]{Action}
Abarbanel, H. D.~I. (2013).
\newblock {\em Predicting the future: completing models of observed complex
  systems}.
\newblock Springer.

\bibitem[Doucet et~al., 2001]{ImportanceSamplingChapter}
Doucet, A., De~Freitas, N., and Gordon, N. (2001).
\newblock An introduction to sequential monte carlo methods.
\newblock In {\em Sequential Monte Carlo methods in practice}, pages 3--14.
  Springer.

\bibitem[Hunt et~al., 2007]{EKF}
Hunt, B.~R., Kostelich, E.~J., and Szunyogh, I. (2007).
\newblock Efficient data assimilation for spatiotemporal chaos: A local
  ensemble transform kalman filter.
\newblock {\em Physica D: Nonlinear Phenomena}, 230(1):112 -- 126.
\newblock Data Assimilation.

\bibitem[Karimi and Paul, 2010]{LChaos1}
Karimi, A. and Paul, M.~R. (2010).
\newblock Extensive chaos in the lorenz-96 model.
\newblock {\em Chaos: An Interdisciplinary Journal of Nonlinear Science},
  20(4):043105.

\bibitem[Kostuk, 2012]{kostuk}
Kostuk, M. (2012).
\newblock Synchronization and statistical methods for the data assimilation of
  hvc neuron models.
\newblock {\em PhD Dissertation in Physics, University of California, San
  Diego}.

\bibitem[Laplace, 1774]{laplaceold}
Laplace, P.~S. (1774).
\newblock Memoir on the probability of causes of events.
\newblock {\em M\'{e}moires de Math\'{e}matique et de Physique, Tome
  Sixi\`{e}me}, pages 621--656.

\bibitem[Laplace, 1986]{laplacenew}
Laplace, P.~S. (1986).
\newblock Memoir on the probability of the causes of events.
\newblock {\em Statistical Science}, 1(3):364--378.
\newblock Translation to English by S. M. Stigler.

\bibitem[Lorenz and Emanuel, 1998]{LChaos2}
Lorenz, E.~N. and Emanuel, K.~A. (1998).
\newblock Optimal sites for supplementary weather observations: Simulation with
  a small model.
\newblock {\em Journal of the Atmospheric Sciences}, 55(3):399--414.

\bibitem[{Metropolis} et~al., 1953]{MC}
{Metropolis}, N., {Rosenbluth}, A.~W., {Rosenbluth}, M.~N., {Teller}, A.~H.,
  and {Teller}, E. (1953).
\newblock {Equation of State Calculations by Fast Computing Machines}.
\newblock {\em J. Chem. Phys.}, 21:1087--1092.

\bibitem[Murty and Kabadi, 1987]{murty87}
Murty, K.~G. and Kabadi, S.~N. (1987).
\newblock Some np-complete problems in quadratic and nonlinear programming.
\newblock {\em Mathematical Programming}, 39:117--129.

\bibitem[Wachter and Biegler, 2006]{ipopt}
Wachter, A. and Biegler, L.~T. (2006).
\newblock On the implementation of a primal-dual interior point filter line
  search algorithm for large-scale nonlinear programming.
\newblock {\em Mathematical Programming}, 106(1):25--57.

\bibitem[Ye et~al., 2014]{ye2014precision}
Ye, J., Kadakia, N., Rozdeba, P.~J., Abarbanel, H. D.~I., and Quinn, J.~C.
  (2014).
\newblock Improved variational methods in statistical data assimilation.
\newblock {\em Nonlinear Processes in Geophysics}, 22(2):205--213.

\bibitem[Ye et~al., 2015a]{ye-et-al}
Ye, J., Kadakia, N., Rozdeba, P.~J., Abarbanel, H. D.~I., and Quinn, J.~C.
  (2015a).
\newblock Improved variational methods in statistical data assimilation.
\newblock {\em Nonlinear Processes in Geophysics}, 22(2):205--213.

\bibitem[Ye et~al., 2015b]{ye2015physrev}
Ye, J., Rey, D., Kadakia, N., Eldridge, M., Morone, U., Rozdeba, P., Abarbanel,
  H. D.~I., and Quinn, J.~C. (2015b).
\newblock A systematic variational method for statistical nonlinear state and
  parameter estimation.
\newblock {\em Physical Review E}.

\bibitem[Zupanski, 2005]{MLEF}
Zupanski, M. (2005).
\newblock Maximum likelihood ensemble filter: Theoretical aspects.
\newblock {\em Monthly Weather Review}, 133(6):1710--1726.

\end{thebibliography}
\bibliographystyle{plain}

\end{document}